\documentclass[preprint,aps,prx,showpacs,superscriptaddress,nofootinbib]{revtex4-1}

\usepackage{amssymb,amsbsy, amsmath}
\usepackage[dvipsnames,usenames]{color}
\usepackage{graphicx}% Include figure files
\usepackage{rotating}
\usepackage{dcolumn}% eqnarray table columns on decimal point
\usepackage[normalem]{ulem} %temporary
\usepackage[caption2]{ccaption}
\usepackage{xr}
\usepackage{xr-hyper} 
\usepackage[draft]{hyperref}% add hypertext capabilities

\usepackage{zref-xr}
\zxrsetup{toltxlabel}

\begin{document}
\title{From the betweenness centrality in street networks to structural invariants in random planar
graphs}
%Structural invariants in planar graphs: betweenness centrality of urban street networks} 

%% List of institution addresses, in command form.
\newcommand{\RochesterP}{Department of Physics \& Astronomy, University of Rochester, Rochester, New York, 14627}
\newcommand{\RochesterC}{Goergen Institute for Data Science, University of Rochester, Rochester, New York, 14627}

\newcommand{\France}{Institut de Physique Th\'eorique, CEA, Gif-sur-Yvette, France}

\newcommand{\france}{Centre d'Analyse et de Math\'ematique Sociales, EHESS, Paris Cedex 6, France}

\newcommand{\floor}[1]{\lfloor #1 \rfloor}

\author{Alec Kirkley}                     \affiliation{\RochesterP}

\author{Hugo Barbosa}                     \affiliation{\RochesterP}

\author{Marc Barthelemy}                     \affiliation{\France}
\affiliation{\france}

\author{Gourab Ghoshal}                     \affiliation{\RochesterP}
 \affiliation{\RochesterC}

\noaffiliation

\begin{abstract}

We demonstrate that the distribution of betweenness centrality (BC), a global structural metric based on network flow, is an invariant quantity in most planar graphs. We confirm this invariance through an empirical analysis of street networks from 97 of the most populous cities worldwide, at scales significantly larger than previous studies. We also find that the BC distribution is robust to major alterations in the network, including significant changes to its topology and edge weight structure, indicating that the only relevant factors shaping the distribution are the number of nodes and edges as well as the constraint of planarity. Through simulations of random planar graph models and analytical calculations on Cayley trees, this invariance is demonstrated to be a consequence of a bimodal regime consisting of an underlying tree structure for high BC nodes, and a low BC regime arising from the presence of loops providing local path alternatives. Furthermore, the high BC nodes display a non-trivial spatial dependence, with increasing spatial correlation as a function of the number of edges, leading them to cluster around the barycenter at large densities. Our results suggest that the spatial distribution of the BC is a more accurate discriminator when comparing patterns across cities. Moreover, the BC being a static predictor of congestion in planar graphs, the observed invariance and spatial dependence has practical implications for infrastructural and biological networks. In particular, for the case of street networks, as long as planarity is conserved, bottlenecks continue to persist, and the effect of planned interventions to alleviate structural congestion will be limited primarily to load redistribution, a feature confirmed by analyzing 200 years of data for central Paris.

\end{abstract}

\maketitle

\date{}

%\section{Introduction}
\noindent {\large{\bf Introduction}}
\vspace{5px}

Recent years have witnessed unprecedented progress in our understanding of spatial networks that are pervasive in biological, technological and infrastructural systems~\cite{bio_planar_nets,Barthelemy:2011dq}. Despite a large number of studies on these objects, in disciplines ranging from mathematics to geography, physics or biology, their structural properties are not fully understood and modeling frameworks remain incomplete. These networks are quite relevant in 
urban systems
~\cite{Bretagnolle_2006, Bettencourt2010, Pan:1fh, Batty:2012dw,barthelemy2016} where analysis of their structural properties has uncovered unique characteristics of individual cities as well as demonstrated surprising statistical commonalities manifested as scale invariant patterns across different urban contexts~\cite{Goh:2016fg, Bettencourt1438, Kalapala_2006}.  Patterns of streets and roads are particularly important, allowing residents to navigate the different functional components of a city.  Different street structures result in varying levels of efficiency, accessibility, and usage of transportation infrastructure~\cite{youn2008, Cardillo:2006fk, Justen:2013kl, Witlox:2007eq, daFCosta:2010er,  Wang:2012jn, Kang:2012by}; consequently structural characteristics of roads have been of great interest in the literature~\cite{Haggett_1969,Lammer_2006,Wang:2011ds,Rui:2013jd, Louf:2014jz, Strano:2013hf,Masucci:2009ja}.

\begin{figure*}[htp]
%\captionsetup{format=plain}
\centering
\includegraphics[width=0.8\linewidth]{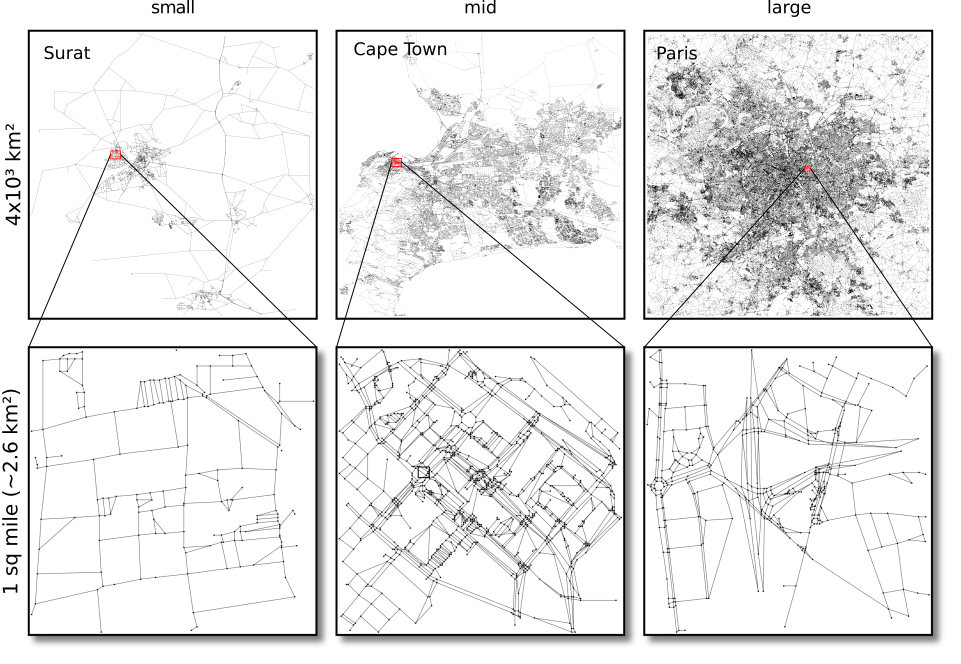}\\
\caption{{\bf Street networks at multiple scales.} Cities split into  three categories based on the number of nodes (intersections) in the sampled street networks: small ($N \sim 10^3$), medium ($N \sim 10^4)$ and large ($N \sim 10^5$). The upper panel shows the networks at the full sampled range of $~10^{3}$ square-kilometers, whereas the lower panel shows selected smaller samples (on the order of one-square-mile).}
\label{fig:fig0}
\end{figure*}

Street networks fall into the category of planar graphs~\cite{Clark_book}, and their edges constitute a \emph{physical} connection, as opposed to relational connections found in many complex networks~\cite{Newman_book}. In particular, the geographical embedding (or spatial constraint) leads to strong effects on network topology with limitations on the number of long-range connections  and the number of edges incident on a single node (its degree $k$)~\cite{Aldous_2013,Aldous_2016}. Degree-based network measures, while well-studied on such systems, lead to rather uninteresting results; for example the degree distribution is strongly peaked, and related metrics such as clustering and assortativity are high~\cite{Barthelemy:2011dq}. Instead, more information can be gleaned from non-local higher-level metrics such as those based on network centralities, which while strongly correlated with degree in non-spatial networks~\cite{Ghoshal_2011}, display non-trivial behavior in planar networks~\cite{crossover}. Among the more studied and illuminating of such metrics is the betweenness centrality (BC), a path-based measure of the importance of a node in terms of the amount of flow passing through it~\cite{freeman}. More precisely, the BC for node $i$ is defined as
\begin{equation}
g_B(i)=\frac{1}{\mathcal{N}} \sum_{s\neq t\in V}\frac{\sigma_{st}(i)}{\sigma_{st}},
\label{eq:bc}
\end{equation}
where $\sigma_{st}$ is the number of shortest paths going from nodes $s$ to $t$ and $\sigma_{st}(i)$ is the number of these paths that go through $i$~\cite{freeman} . Here $\mathcal{N}$ is a normalization constant, typically of order $N^2$ where $N$ is the number of nodes, although for reasons that will be apparent later in the manuscript, we will use here the unnormalized version ($\mathcal{N}=1$). 

In principle, one can define a variety of different shortest paths, the number of hops (in the purely topological case), the shortest distance between two points (geodesic) if the edges are weighted according to Euclidean distances, taking into account route preferences if edges are weighted according to a ``cost function" such as capacity or speed-limits, or indeed some combination of the above. Incorporating this structural information into the edge-weights, the BC can be used as a proxy for predicted traffic flow~\cite{holme,Ashton:2005ck,Jarrett:2006ha}. In such a setting the paths can be considered as the optimal routes between locations, and thus nodes with high BC should expect to receive more traffic. Consequently, in what is to follow, we will focus on the \emph{weighted} node BC, calculated using the standard Brandes algorithm~\cite{Brandes_2001}.

A number of studies have been conducted on the BC in planar graphs~\cite{Roswall_2005,Jiang:2007ch,Chan:2011bsa} finding among other things, a complicated spatial behavior of the high BC nodes~\cite{Lammer_2006,Lion_2017}, and in the case of street networks, connections to the organization and evolution of cities~\cite{Crucitti:2006haa, porta1, Barthelemy_2013}. For non-planar graphs the average BC scales with the degree $k$ in a power law fashion thus $g_B(k)=\sum_{i\vert k_i=k}\frac{g_B(i)}{N(k)}\propto k^{\eta}$,  where $N(k)$ is the number of nodes of degree $k$, and $\eta$ is an exponent depending on the graph~\cite{barthbet}. In planar graphs, however, the BC behaves in a more complex manner, as now both topological \textit{and} spatial effects are at play. While for a regular lattice, the BC is a function of the distance from the barycenter, with increasing disorder (as found in real planar networks) the BC will in general be a function of the local topology as well as the distance from the barycenter~\cite{Lion_2017}. 

Given their practical relevance as well as the relative abundance of data, street networks have proven to be an excellent platform on which to study the properties of planar graphs. Existing analyses of the BC in particular, however, suffers from limitations of scale (unlike other structural properties,  see \cite{strano2017scaling} for a recent global description), and most comparative studies of the BC across streets in cities are typically restricted to one-square-mile samples, while studies on more extensive street-maps have been examined for at most tens of cities limited to those in Europe or North America~\cite{porta1, Crucitti:2006haa, Cardillo:2006fk, Lion_2017, Barthelemy_2013}. Furthermore, there have been limited studies of the BC distribution in its entirety, with the majority of analyses instead focusing on the average  BC (proportional to the average shortest path \cite{gago2012notes}) or on its maximum value~\cite{narayan2011large,jonckheere2011euclidean}.

To fill this gap in our understanding of this important class of networks, we conduct here a large-scale empirical study of the BC across 97 of the world's largest cities as measured by population (details on dataset in Materials and Methods, Supplementary Note \ref{sisec:data} and Table \ref{sitab:networkproperties}). The cities are sampled from all six inhabited continents and the analysis is conducted at a scale of the order of three thousand square-kilometers. We group the different cities in three main categories according to their size (Fig.~\ref{fig:fig0}),
from small ($N\sim 10^3$ nodes), medium ($N\sim 10^4$) to large road networks ($N\sim 10^5$).

%While our findings are of relevance to physical infrastructure, urban planning and public policy, their reach extends beyond urban systems, being applicable to planar networks in general, with potential implications for biological and ecological systems~\cite{urban2009graph}.

%We demonstrate that this change of scale leads to some rather remarkable findings, among them that the BC distribution across cities turns out to be relatively %invariant after rescaling with the number of street intersections $N$ and accounting for finite size-effects. 

%Despite the apparent (and) obvious differences stemming from spatial layout, topological\hugo{,} and geographic constraints, the high BC nodes across all %cities lie on an underlying tree structure, in addition to displaying a highly non-trivial spatial dependence, with a pattern that transitions from topological to %spatial, as a function of the number of roads. Analysis of historical data covering the central portion of Paris, spanning two hundred years,  indicates that these %features are robust to local topological changes in the city. 

%\section{Betweenness at different scales and rescaling}  
\vspace{5px}
\noindent {\large{\bf Results}}
\vspace{5px}

\noindent {\bf Betweenness at different scales and rescaling}  

In Fig.~\ref{sifig:bcatscale}A we show the betweenness probability distribution for a selection of the three categories of cities at the resolution of two and a half square kilometers (or one-square-mile), plotted in a log-linear scale. One sees significant variability between cities, within and across categories, with mostly exponential tails (Fig.~\ref{sifig:exponential_fits}), as also seen for similar samples in~\cite{Crucitti:2006haa,porta1}. This is somewhat expected given the small sample size, as even controlling for the number of street intersections, the topology of cities are different due to geographic and spatial constraints~\cite{Lee_2017,Clark:2016vj}. Indeed, these variations may show up \emph{within the same city} where multiple samples of a similar resolution within a city display important fluctuations (Fig.~\ref{sifig:bcatscale}B). In all cases, we observe a range of behavior in the tails of the BC ranging from peaked to broad distributions, reflecting local variation in the street network structure and fluctuations in the data. One begins to see a dramatic difference when increasing the scale to three thousand square-kilometers  where the distribution in all cities start to look similar (Fig.~\ref{sifig:bcatscale}C,D). We observe that the BC distribution for cities within each category is virtually \emph{identical}, and also that the distribution is \emph{bimodal}, with two regimes separated by a bump roughly at  $g_B \sim N$. For larger values of the BC we observe a slow decay signaling a broad distribution. 

%The combination of these features appear to have been overlooked or missed in existing work primarily due to the low resolution of the sampled street networks, or excess noise due to linear binning on a logarithmic $x$-axis~\cite{bet_weighted}. 

\begin{figure*}[htp]
\centering
\includegraphics[width=0.8\linewidth]{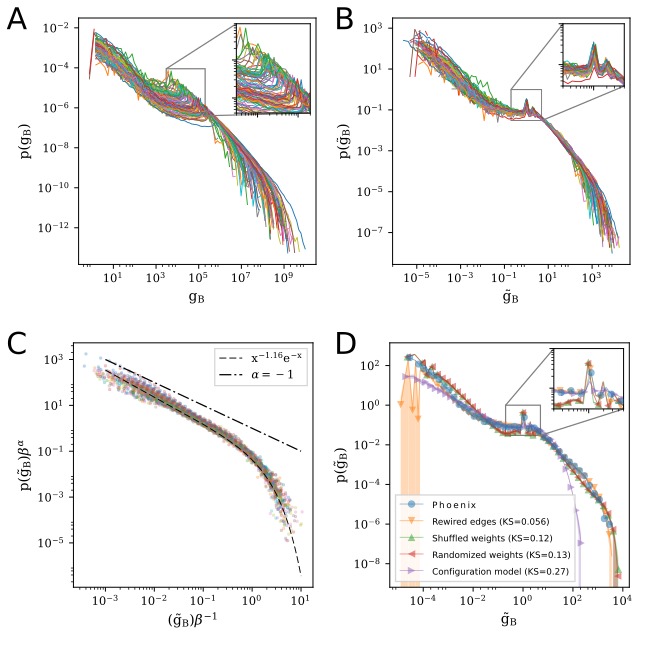}
\caption{{\bf Betweenness invariance in urban streets.} {\bf (A)} The betweenness pdf for all 97 cities at full-resolution. The peak of the distribution for each city is shown as inset. {\bf (B)} The version of the distribution after rescaling by the number of nodes $N$ showing the alignment of the peaks across all cities (also shown as inset). {\bf (C)} The collapse of the tails after rescaling with respect to $\beta$. The dashed line shows the analytically computed asymptotic scaling for a Cayley-tree (Eq.~\ref{eq:cayley}). {\bf (D)} The BC distribution of various random graph models described in the text compared to the baseline distribution of Phoenix as a representative example. Shaded area reflects fluctuations around the average over hundred realizations of each model. Apart from the (non-spatial) configuration model, we see minimal changes in the location of the peak(s) (zoomed in inset) or shape of the tail. Also shown are 2-sample KS statistics for phoenix and its corresponding random graph models (details in Figs.~\ref{sifig:KS_table} and~\ref{sifig:P_table}).}
\label{fig:fig2}
\end{figure*}

These trends are apparent in the BC distribution across \emph{all} 97 cities in our data as seen in Fig.~\ref{fig:fig2}A with the two regimes being separated by bumps spread across an interval of $10^3 \leq g_B \leq 10^5$ corresponding to the range of $N$ in our data. Indeed rescaling the betweenness of each node by the number of vertices in the network $g_B\to \tilde{g_B}=g_B/N$, we see the distributions collapse on a single curve with a unique bump separating two clear regimes as seen in~Fig.\ref{fig:fig2}B, although some variability exists resulting from differences in the number of edges. Fitting the distribution of $\tilde g_B = g_B/N$ with the function
\begin{equation}
p(\tilde g_B) \sim \tilde g_B^{~-\alpha} e^{-\tilde g_B/\beta},
\label{eq:fit}
\end{equation}
results in a tightly bound range for $\alpha \approx 1$ and a broad size-dependent distribution for $\beta$ (Fig.~\ref{sifig:betweennessalphabetawithbetafitsi}). Rescaling the tail with respect to $\beta$ results in a collapse of the curves for all cities as seen in Fig.~\ref{fig:fig2}C. (Details of the rescaling and fitting procedures in Supplementary Note~\ref{sisec:fitting}, Fig.~\ref{sifig:normalizedbetweennessdistributionall} and Tab. \ref{sitab:fit_pars}). 

%\section{Determinants of the BC distribution} 
\vspace{5px}
\noindent {\bf Determinants of the BC distribution}
\vspace{5px}

Given that cities are ostensibly quite different in terms of geography or space, as well as  their levels of infrastructure and socioeconomic development, the observed invariance is quite striking. To investigate the factors behind this behavior, we next systematically probe  the effect of the main features that may be influencing the BC distribution. Examining Eq.~\eqref{eq:bc}, apart from its obvious dependence on the number on nodes $N$ and the number of edges $e$, the other primary factors are the {\bf (a)} Local topology---the local connectivity patterns of a street intersection as governed by its degree distribution; {\bf (b)} Distribution of edge weights that can correspond either to Euclidean distances or some scalar quantity such as speed-limits or capacity; and {\bf (c)} Planarity---the effect of space. To do so we select the BC distribution of a number of cities as baseline and generate multiple variants of random graphs to compare with the original. In Fig.~\ref{fig:fig2}D we show Phoenix (blue circles) as a representative example of a city on which we perform this analysis.

%Note that for all variants of the random graph models, we plot the distribution of $\tilde g_B$. Additionally, for comparison purposes, we compute the %2-sample KS statistic, the details of which can be found in Supplementary figures \ref{sifig:KS_table} and \ref{sifig:P_table}.

%\subsection{Effect of local topology}
\vspace{5px}
\noindent{\it Effect of local topology}

In principle, the BC of a node can be rather sensitive to local changes in topology. Consider, for example, the case of a ``bridge node" that connects two disjoint clusters via connections to a single node in each cluster. Such a node has a high BC as it necessarily lies between \emph{all} shortest paths between the two clusters. Yet, simply by placing an edge directly between the clusters, one can dramatically decrease the BC of the bridge node. To investigate such effects---which amounts to varying the local neighborhood of a given street intersection---we fix the spatial position of nodes on the 2D plane  and generate a Delaunay Triangulation (DT)~\cite{lee1980two} of the street network. The DT corresponds to the maximum number of edges that can be laid down between a fixed number of nodes distributed within a fixed space, without any edge-crossings. Edges are then randomly eliminated until their number corresponds exactly to our baseline example of Phoenix. A hundred realizations of this procedure was conducted, having the effect of rewiring the local neighborhood of intersections---by changing a node's degree and its neighbors---while still maintaining planarity. In Fig.~\ref{fig:fig2}D we plot the average of these realizations (orange triangles), showing differences with the original street network in the lower range of the distribution, yet showing minimal change in both the location of the peak as well as the tail of the distribution. Similar random graphs were generated using a number of other cities as baseline showing the same behavior (Fig.~\ref{sifig:nullmodels_all}). 

%\subsection{Effect of edge weights} 
\vspace{5px}
\noindent{\it Effect of edge weights}

Next we investigate the effect of Euclidean distances on the BC distribution.  We fix the number of nodes $N$ and instead of fixing their positions according to the empirical pattern, we now distribute them {\it uniformly in the 2D plane} with a scale determined by the spatial extent of the city considered. Having done this, we generate the DT of the street network and randomly remove edges until we match the number of roads in the data. A hundred different realizations of this procedure has the effect of stretching/compressing the city in multiple directions (either dispersing high density areas or compressing very long road segments) and therefore generating a distribution of distances that are markedly different from the original (Fig.~\ref{sifig:cityweights}). Fig.~\ref{fig:fig2}D (red triangles) suggests that while this has a marginally stronger effect than edge rewiring, the tails of the original and perturbed distributions are quite similar within the bounds of the error-bars. Furthermore, the positions of the peaks remain unchanged. Varying the area (and therefore density of nodes) and conducting the same procedure over multiple cities yielded identical results (Fig.~\ref{sifig:dtscales}), suggesting that the distribution of (spatial) edge-weights has negligible effect on the BC distribution.

While the procedure outlined above does not preserve the local topology (degree of individual nodes) it is possible to change the edge-weights while preserving the degree sequence of nodes. This can be done by taking the original street network and randomly sampling from its associated distribution of distances, assigning each edge a number from this distribution---the edge-weights now \emph{do not} correspond to physical distances but can be interpreted instead as a cost function such as speed-limits, travel demand, or road capacity. In  Fig.~\ref{fig:fig2}D we show the average of this process over a hundred realizations (green triangles) where each realization corresponds to a reshuffling of the edge weights over the network. We now begin to see some changes in the distribution with a minor shift in the position of the peaks and a moderately heavier tail, although no drastic modifications are apparent. Strikingly, sampling from a whole family of distributions for the edge weights (exponential, power-law, log-normal) produced identical results (Fig.~\ref{sifig:otherdists}), indicating little-to-no dependence on the specific nature of the weights (spatial or non-spatial).

%In other words, while there was a small deviation from the original BC distribution as a result of decoupling the edge-weights from spatial dependence, comparing the resulting perturbed distributions yielded little-to-no dependence on the specific distribution of edge-weights.  

%\subsection{Relaxing Planarity}
\vspace{5px}
\noindent{\it Relaxing planarity}

Finally, we probe the effects of relaxing the condition of planarity. Fixing $N$, the degree-sequence, and assigning weights sampled from the distance distribution of Phoenix, we use the configuration model~\cite{Newman_2001} (given a degree sequence, a random graph is constructed by uniformly and randomly choosing a matching on the degree stubs emanating from each node) to generate one hundred non-spatial versions of the street network resulting in the markedly different curve in Fig.~\ref{fig:fig2}D (purple triangles). The shape of the curve is in line with the known dependence of $g_B$ on the degree for non-spatial networks, with a distribution of degrees peaked around $k=3$ (Figs. \ref{sifig:degreedistribution} and \ref{sifig:degreedistributionall}). The markedly different shape of the curve as compared to the actual street network shows that planarity appears to be the dominant factor specifying the BC distribution, with topological effects and edge-weights playing only a negligible role. While this provides an explanation for the observed similarity across cities despite their significant geospatial variations, it does not by itself provide an explanation for the form of the distribution, its scaling with $N$, nor its bimodality, and we will provide in the following some theoretical arguments. 

%\begin{figure*}[htp]
%\centering
%\includegraphics[width=0.5\linewidth]{man_figs/Fig3}
%\caption{{\bf Effect of edge-density $\rho_e$ on the betweenness:} $N =10^{4}$ nodes were randomly distributed on the 2D plane and their DT was generated. Edges were removed until the desired edge-density $\rho_e$ was reached. The left panel shows the averaging over a hundred realizations of the resulting BC distribution ranging from the MST constructed over all nodes {\bf (A)} to the DT {\bf (D)} with increasing $\rho_e$. The orange shaded area corresponds to fluctuations around the average of the realizations, while the silver and white shades separate the ``tree-like" region from the ``loop-region" respectively. The right panel shows a single instance of the actual generated network corresponding to each $\rho_e$. Shown in red are the nodes in the 90'th percentile and above in terms of their BC value.}
%\label{fig:fig3}
%\end{figure*}

\begin{figure*}[t!]
\includegraphics[width=0.75\linewidth]{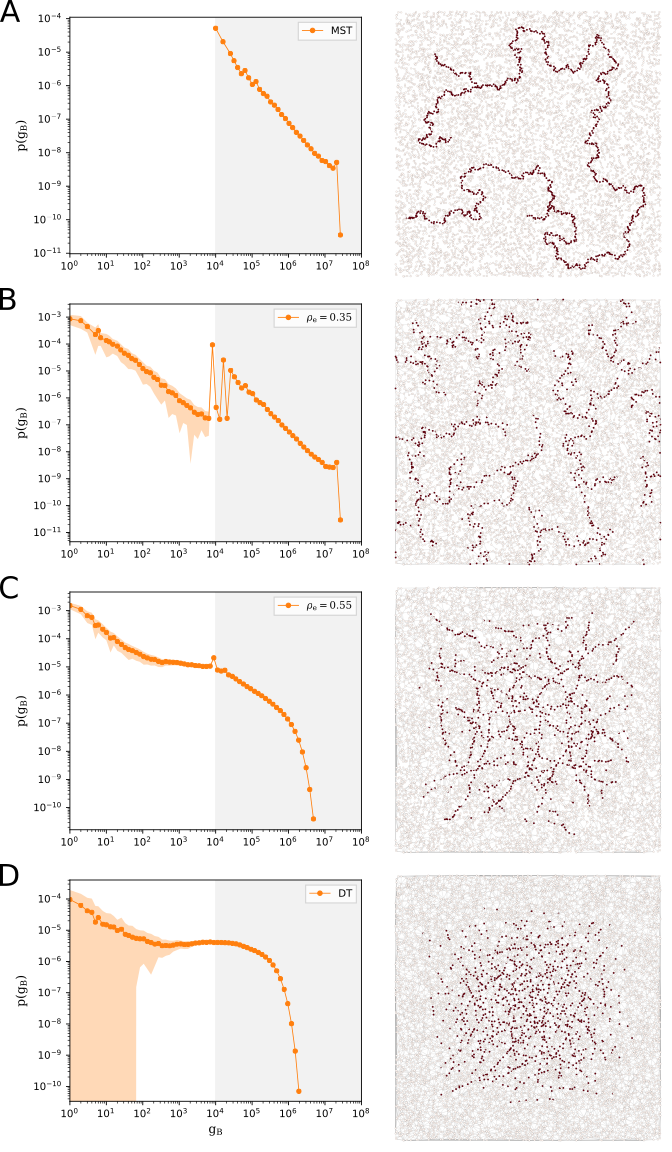}
\caption{{\bf Effect of edge-density $\rho_e$ on the betweenness}}
\label{fig:fig3}
\end{figure*}

\begin{figure}[t]
\contcaption{ $N =10^{4}$ nodes were randomly distributed on the 2D plane and their DT was generated. Edges were removed until the desired edge-density $\rho_e$ was reached. The left panel shows the averaging over a hundred realizations of the resulting BC distribution ranging from the MST constructed over all nodes {\bf (A)} to the DT {\bf (D)} with increasing $\rho_e$. The orange shaded area corresponds to fluctuations around the average of the realizations, while the silver and white shades separate the ``tree-like" region from the ``loop-region" respectively. The right panel shows a single instance of the actual generated network corresponding to each $\rho_e$. Shown in red are the nodes in the 90'th percentile and above in terms of their BC value.}
\end{figure}

%\section{Modeling the BC distribution}
\vspace{5px}
\noindent {\bf Modeling the BC distribution} 

A clue for the bimodal behavior stems from the fact that it is peaked at $N$, a feature reminiscent of nodes adjacent to the leaves of a \emph{minimum spanning tree} (MST).  The MST consists of the subset of edges connecting all nodes with the minimum sum of edge-weights~\cite{Graham_1985} and whose betweenness value is of $O(N)$, specifically $N-2$ for degree two nodes adjacent to leaves. Indeed all paths from the leaf to $N-2$ other nodes have to go through this node. In the context of street networks, their analogs are nodes adjacent to dead-ends. An examination of the BC distribution of trees therefore, may provide an explanation for the scaling behavior found in our data. 

%\subsection{Cayley Tree Approximation}
\vspace{5px}
\noindent{\it Cayley tree approximation}

While deriving an exact analytical expression for the BC distribution of generalized MST's is challenging, one can make progress by approximating it as a $k$-ary tree (where each node has a branching ratio bounded by $k$). Given that the degree distribution of streets is tightly peaked (Fig. \ref{sifig:degreedistribution}), we can make a further approximation by assuming a fixed branching ratio, in which case the $k$-ary tree reduces to a Cayley tree where all non-leaf nodes have degree $k$. Assuming all leaf nodes are at the same depth $l$ and adopting the convention $l=L$ for the leaf level and $l = 0$ for the root, a simple calculation reveals that for a node $v$ at level $l$, the betweenness scales as $g_B(v\vert k,l)\sim O(Nk^{L-l})$. After a sequence of manipulations (Materials and Methods), it can be shown that
\begin{equation}
P(g_B) \approx \frac{k^{\log_k\left (\frac{AN}{g_B}\right)}}{N}=Ag_B^{-1},
\label{eq:cayley}
\end{equation}
indicating that the node betweenness of a Cayley tree scales with exponent $\alpha = 1$, consistent with previous calculations of the link betweenness~\cite{szabo2002shortest}. This provides a possible explanation for both the scaling with $N$ as well as the form of the tail found in the empirical measurements of the BC of city streets (Eq.~\eqref{eq:fit}). Indeed, this implies an underlying tree structure on which the high BC nodes of all cities lie, indicating that the majority of flow is concentrated around \emph{a spanning tree} of the street network~\cite{wu2006transport}. While a similar feature is seen for the BC of weighted (non-planar) random graphs, this is only true for specific families of weight distributions~\cite{bet_weighted}, a factor that has little-to-no effect in planar graphs.

%\subsection{A simple model with variable density}
\vspace{5px}
\noindent{\it A simple model with variable density}

Of course, street networks are not trees and contain loops given by the cyclomatic number $\Gamma=e-N+1$ (for a connected component) where $N$ is the number of nodes and $e$ is the number of edges. In the absence of any loops (such as in the MST) we have that $N =e+1$, and given that $N$ is fixed, the addition of any further edges will necessarily produce loops leading to alternate local paths for navigation. With an increasing number of edges (and therefore more alternate paths), one would expect a large fraction of the (previously) high BC nodes lying on the MST to be bypassed, decreasing their contribution to the number of shortest paths. This will induce the emergence of a low BC regime as well as increasingly sharp cutoffs in the tail, in line with the empirical observations of street networks (Fig.~\ref{fig:fig2}).

In order to study the impact of increasing edges on the BC distribution, we study a simple model of random planar graphs. Given that $e \sim O(N)$ and that $N$ itself varies over three orders of magnitude in our dataset, we define a control parameter which we call the \emph{edge density} thus, 
\begin{equation}\label{eqn:edge_density}
\rho_e=\frac{e}{ e_{DT}},
\end{equation}
defined as the fraction of extant edges $e$ compared to the maximal number of possible edges constructed on the set of nodes (given by intersections), as determined by its Delaunay Triangulation $e_{DT}$, and which varies from $\rho_e\approx 1/3$ for the MST to $\rho_e=1$ for the DT~\cite{lee1980two}. For a maximally planar graph (i.e one in which no more edges can be added without violating the planarity constraint), we have that $e_{DT} \approx 3N$, so the metric captures the ratio of edges to nodes, or in the context of street networks, the average degree $\langle k \rangle$ of street intersections. Of course in the latter case, the limit $\rho_e \rightarrow 1$ is unlikely given some of the geographic constraints inherent in cities. 

Having defined this control parameter, we distribute $N$ nodes uniformly in the 2D plane and we first study the MST. In order to be able to vary the density, we generate the Delaunay triangulation on the set of nodes and remove edges until we reach the desired value for $\rho_e$. The left panel of Fig.~\ref{fig:fig3} shows the  BC distribution resulting from a hundred realizations of this procedure for $N =10^{4}$ and for increasing values of $\rho_e$ from the MST (top) to the DT (bottom). The BC distribution for the MST seen in Fig.~\ref{fig:fig3}A is peaked at $N$ and is bounded by $N^2/2$ which gives here a range of order $[10^4, 10^8]$. In this interval, the BC distribution follows a form close to our calculation for the Cayley tree~(Eq.\ref{eq:cayley}). As one increases $\rho_e$ and creates loops in the graph, we see the emergence of a bimodal form, with a low BC regime resulting from the bypassing of some of the high BC nodes due to the presence of alternate paths (Fig.~\ref{fig:fig3}B). Note that the distribution continues to be peaked at $N$ and the tail maintains its shape. As $\rho_e$ is further increased, the distribution gets progressively more homogeneous, yet remains peaked around $N$ even as we approach the limiting case of the DT (Fig.~\ref{fig:fig3}D). As a guide to the eye, we shade the ``tree-like" region from the ``loop-like" region separated by the peak at $N$. 

These results suggest that the observed bimodality seen in the BC distribution for cities, stems from the presence of a backbone of high BC nodes belonging to the MST, decorated with loops. Nodes on these loops contribute to the low BC regime. The transition between the two regimes---low versus high BC nodes---is determined by the minimum non-zero betweenness value for the MST, which is $O(N)$ and the tail may have different peaks,  determined by the distribution of branches emanating from the tree. Progressively decorating the tree with loops leads to arbitrarily low betweenness values due to the creation of multiple alternate paths, thus smoothing out the distribution, as the betweenness transitions from an interval $[N,N^2/2]$ for the MST to a continuous distribution over $[1,N^2]$ for the DT. 

%\section{Spatial distribution of high BC nodes: characterization}
\vspace{5px}
\noindent{\bf Spatial distribution of high BC nodes: characterization} 

%\subsection{Random planar graphs}

\noindent {\it Random planar graphs} 

The right-hand panel of Fig.~\ref{fig:fig3} shows a single instance of the actual network generated by our procedure for each corresponding edge-density. Highlighted in red are nodes lying in the $90^{th}$ percentile and above in terms of their BC. For these nodes, there is a distinct change in spatial pattern with increasing $\rho_e$. At the level of the MST, they span the network and are tree-like with no apparent spatial correlation. As the network gets more dense, one sees a tendency of these nodes to cluster together and move closer to the barycenter, suggesting a transition between a ``topological regime" and a ``spatial regime". 

\begin{figure*}[t!]
\centering 
\includegraphics[width=0.8\linewidth]{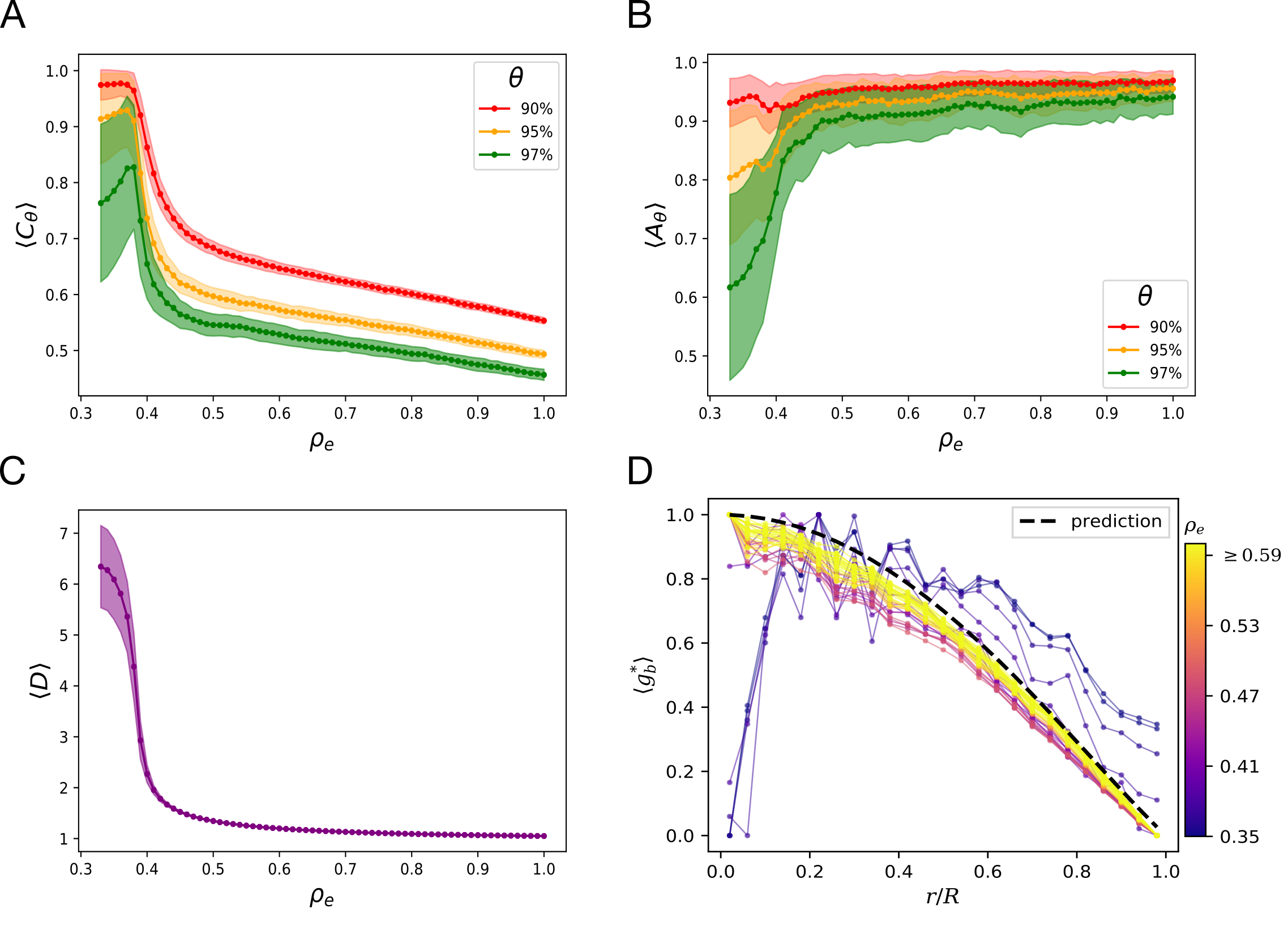} 
\caption{{\bf Quantifying the spatial effect of edge-density $\rho_e$ on high BC nodes} \textbf{(A)} The metric $\langle C_\theta \rangle$ (Eq.~\ref{eq:cluster}) decreases for denser networks, capturing the tendency of the nodes to be increasingly clustered around their center of mass. \textbf{(B)} Correspondingly they also become more isotropic around this center as $A_\theta$ (Eq.~\ref{eq:iso}) approaches $1$. \textbf{(C)} The network also becomes increasingly geometric as indicated by the decrease in the average detour factor $\langle D\rangle$ (Eq.~\ref{eq:detour}) measured for the \emph{full network}, which experiences an abrupt transition around $\rho_e\sim 0.4$.  The shaded regions represent fluctuations over hundred realizations of the randomization procedure. \textbf{(D)} The average BC for nodes at a distance $r$ from the barycenter (rescaled to the interval [0,1]), measured in units of $r/R$ where $R=50$ is the grid boundary. Curves are colored according to the value of $\rho_e$. The dashed line corresponds to the analytical calculation for an infinitely dense random geometric graph~\cite{Giles_2015}.  The metrics are computed for the networks generated in Fig.~\ref{fig:fig3}.}
\label{fig:3_clustering_metrics}
\end{figure*}

To quantify these observed changes, we investigate the behavior of the high BC nodes at percentile $\theta$ through a set of metrics: the clustering $C_\theta$ which measures the spread of high BC nodes around their center of mass, the anisotropy factor $A_\theta$ which characterizes the spatial anisotropy of this set of nodes, and finally, the detour factor $D$ which measures the average extent to which paths between two locations deviate from their geodesic distance. (Details on metrics shown in Materials and Methods)

In Fig.~\ref{fig:3_clustering_metrics}A  we plot the quantity $\langle C_{\theta} \rangle$ for $\theta = 90$, $95$, and $97$ for different values of the edge-density, finding a clear asymptotic decrease with $\rho_e$ (here $\langle \ldots \rangle$ indicates averaging over realizations). Indeed the decrease is approximately by a factor of two from the MST to the DT, confirming the spatial clustering of the nodes to be a robust effect.  In Fig.~\ref{fig:3_clustering_metrics}B the plot of $\langle A_{\theta}\rangle$ in function of $\rho_e$, for the same set of thresholds as before, indicates a growing isotropic layout and is indicative of a transition from a quasi one-dimensional to a two-dimensional spatial regime. This is confirmed by the corresponding decrease in the detour factor shown in Fig.~\ref{fig:3_clustering_metrics}C, where there is a rapid drop around $\rho_e\approx 0.4$ (or equivalently $\langle k\rangle\approx 2$) which is near the density region when the network transitions from a tree-like to a loop-like region. The appearance of loops in the graph has the additional effect of significantly lowering the detour leading to short paths that are increasingly straight in the geometric sense. 

Furthermore, plotting the rescaled average BC of nodes as a function of the distance $r$ from the barycenter (Materials and Methods), shows a monotonic decrease with distance in the high density regime (Fig.~\ref{fig:3_clustering_metrics}D).  For low values of $\rho_e$ there appears no distance dependence of the nodes, whereas for $\rho_e > 0.4$, a clear $r$ dependence emerges with the curves converging to the form seen for maximally dense random geometric graphs as calculated in~\cite{Giles_2015}. (Note that while both planar and geometric graphs are embedded in space, the latter allows for edge-crossings and therefore broader degree distributions and larger number of edges for the same $N$. In light of this difference, the similarity between the two ostensibly different classes of graphs is notable.)   In combination, the structural metrics suggest that while the spatial position of a node is decoupled from its BC value in sparse networks, a strong correlation emerges for increasingly dense networks.

%\begin{figure*}[htp]	
%\centering
%\includegraphics[width=0.8\linewidth]{man_figs/cities_fig_v3}
%\caption{{\bf Spatial behavior of high BC nodes in real cities:} {\bf (A)} Distribution of edge-densities for the 97 cities lie in a narrow range $0.4 \leq \rho_e \leq 0.6$ with most cities peaked at $\rho_e \approx 0.5$. {\bf (B)} Variation of the spatial clustering $\langle C_{\theta} \rangle$ and {\bf (C)} anisotropy ratio $\langle A_\theta\rangle$ with $\rho_e$ for the same range of thresholds used for the random graph models . {\bf (D)} Detour factor for the full street network across cities plotted according to their edge-density. Points are averages over cities within a bin-size of $\rho_e = 0.02$ and the shaded areas represent the fluctuations within the bins. {\bf (E--H)} Spatial layout of intersections in four representative cities in increasing order of $\rho_e$. The color scale goes from purple to yellow with increased BC. The functional trends of the metrics and the geospatial patterns for the cities are consistent with what is observed for the random graph model described in the text.}
%\label{fig:fig4}
%\end{figure*}

\begin{figure*}[t!]	
\centering
\includegraphics[width=\linewidth]{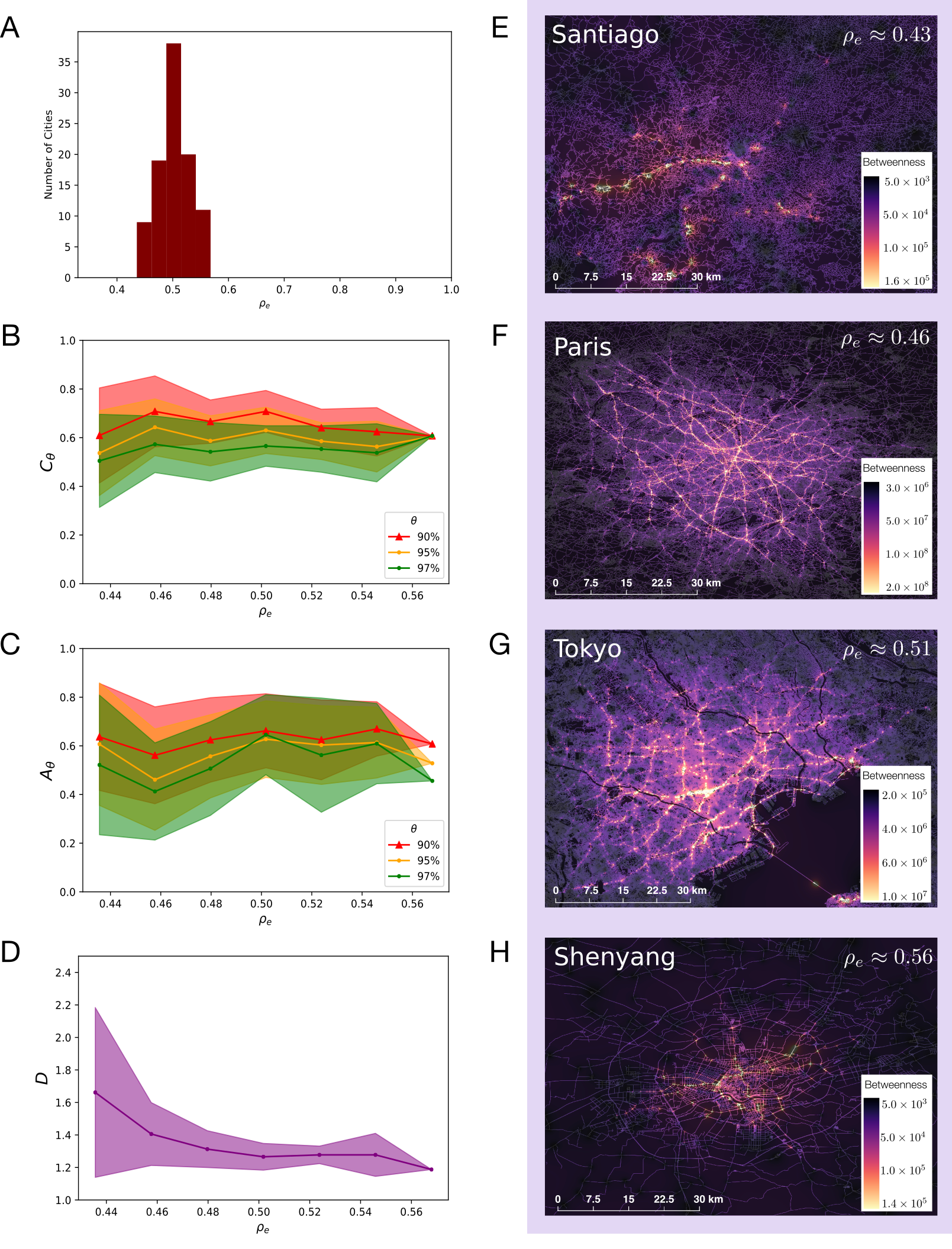}
\caption{{\bf Spatial behavior of high BC nodes in cities}}
\label{fig:fig4}
\end{figure*}

\begin{figure}[t]
\contcaption{ {\bf (A)} Distribution of edge-densities for the 97 cities lie in a narrow range $0.4 \leq \rho_e \leq 0.6$ with most cities peaked at $\rho_e \approx 0.5$. {\bf (B)} Variation of the spatial clustering $\langle C_{\theta} \rangle$ and {\bf (C)} anisotropy ratio $\langle A_\theta\rangle$ with $\rho_e$ for the same range of thresholds used for the random graph models . {\bf (D)} Detour factor for the full street network across cities plotted according to their edge-density. Points are averages over cities within a bin-size of $\rho_e = 0.02$ and the shaded areas represent the fluctuations within the bins. {\bf (E--H)} Spatial layout of intersections in four representative cities in increasing order of $\rho_e$. The color scale goes from purple to yellow with increased BC. The functional trends of the metrics and the geospatial patterns for the cities are consistent with what is observed for the random graph model described in the text.}
\end{figure}

%\subsection{Empirical results}
\vspace{5px}
\noindent {\it Empirical results}

Having observed the spatial behavior of the high BC backbone in random graphs, we next investigate this in the 97 cities. The distribution of $\rho_e$ in Fig.~\ref{fig:fig4}A lies in a tight range ($0.4 \leq \rho_e \leq 0.6$) with the majority of cities peaked at $\rho_e \approx 0.5$. The absence of cities with large edge-densities is to be expected, given the nature of street networks, where a node does not exist independently (as in the random graph) but necessarily corresponds to the intersection of streets. The observed range is notable, as  for one it corresponds to a range of edge densities where a clear bimodal regime exists as seen in Fig.~\ref{fig:fig3}, while the peaked nature of $\rho_e$ provide a further explanation for the observed similarity in BC distributions, given that it is the key controlling parameter. On the other hand, this provides a limited window for checking the spatial trends; indeed the curves for $\langle C_\theta \rangle$, $\langle A_\theta \rangle$ and $D$ shown in Figs.~\ref{fig:fig4}B,C,D are noisy.  Fluctuations arise due to a combination of  smaller samples compared to those generated in our random graph simulations, as well as the averaging over cities with the same edge-density but different $N$. Yet, within the extent of fluctuations, the trend is reasonably consistent with that seen in Fig.~\ref{fig:3_clustering_metrics} for the same range of $\rho_e$.  A clearer picture emerges when looking at individual cities; in Fig.~\ref{fig:fig4}E-H we show the geo-spatial layout of the BC distribution for the full street network in four representative cities arranged in increasing order of edge-density. Santiago, being a city with relatively sparse number of streets, shows a tree-like anisotropic pattern for the high BC nodes that are spread mostly along a single axis of the city. Paris and Tokyo, being in the intermediate range, show a complicated lattice-like structure with loops spanning the spatial extent of the cities. Finally, Shenyang, being a city from the upper range of densities, shows a clear (relatively symmetric) clustering of the high BC nodes around the city center. 

\begin{figure*}[htp]
\centering
\includegraphics[width=0.8\linewidth]{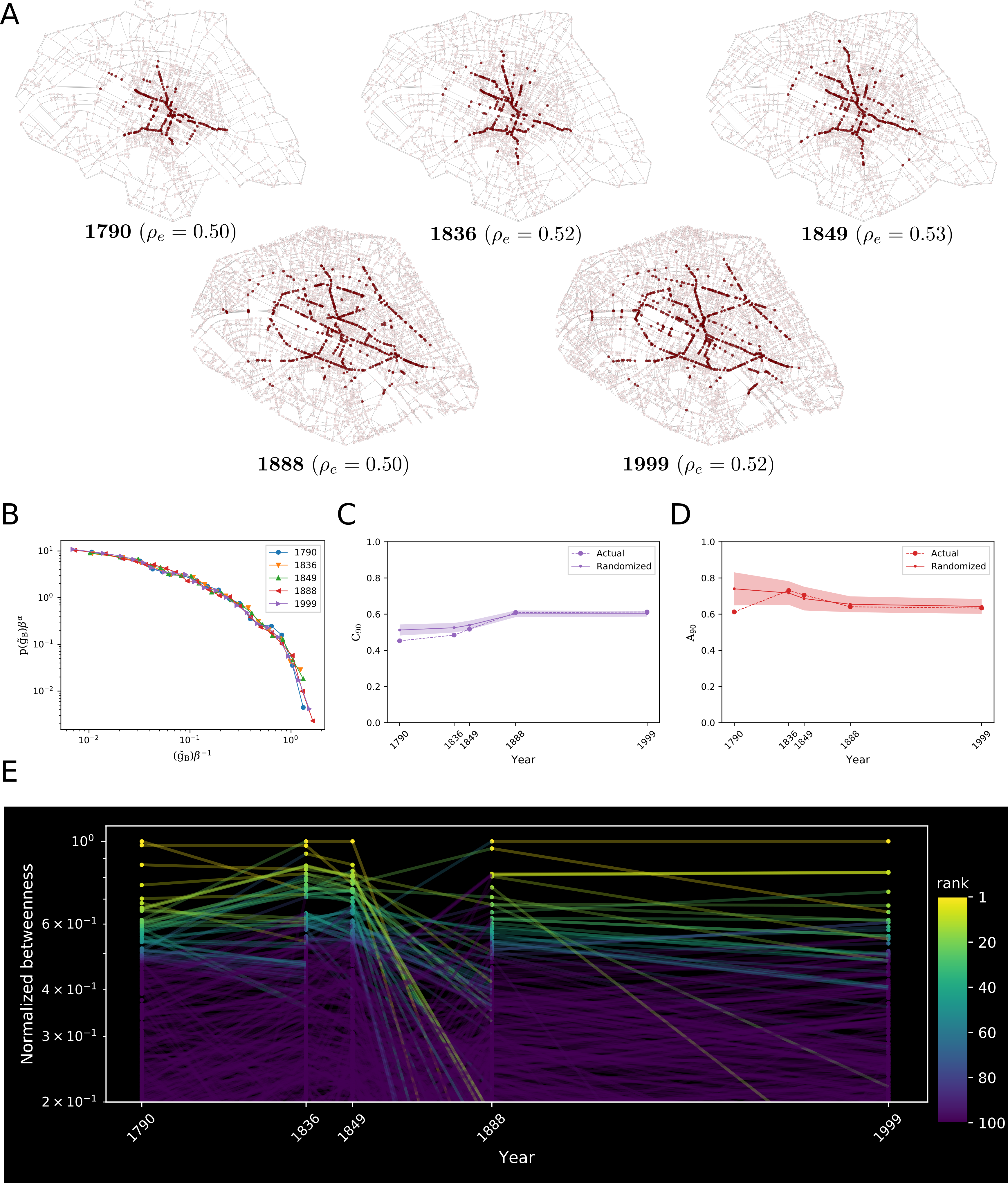}
\caption{{\bf Evolution of central Paris from 1790--1999} {\bf (A)} Five snapshots of a portion of central Paris spanning two hundred years. Colored in red are the nodes corresponding to the 90'th percentile in terms of their BC. {\bf (B)} The rescaled BC distributions (using the same method as in Fig.~\ref{fig:fig2}C) for all five networks showing that they are identical ($\alpha = 1)$.  {\bf (C)} The clustering and {\bf (D)} anisotropy metrics for the nodes in the $90^{th}$ percentile. Also shown are the corresponding metrics for hundred realizations of randomized versions of the networks according to the procedure used in Fig.~\ref{fig:fig3}. {\bf (E)} Temporal evolution of the BC of individual nodes that are present in all five networks. Points are colored according to rank based on BC and lines drawn with reduced opacity for improved visualization.}
\label{fig:paris_evo}
 \end{figure*}

%\section{Beyond static structure: Temporal evolution of BC in cities}

\noindent{\bf Beyond static structure: Temporal evolution of BC in cities} 

The changes in the structure of the random graph shown in Fig.~\ref{fig:fig3}, can be thought of as a proxy for the evolution of a city as it experiences refinements in infrastructure with increased connectivity. While historical data of the evolution of the full street networks in cities is limited, progress can be made by examining smaller subsets. To this effect, we make use of data of five historical snapshots of a portion of central Paris spanning two hundred years (1790--1999), previously gathered to study the effects of central planning by city authorities~\cite{Barthelemy_2013}. The selected portion of Paris is around thirty square kilometers with about $10^3$ intersections and road-segments, and represents the essential part of the city around 1790. This particular period was chosen to examine the effects of the so-called ``Hausmann transformation", a major historical example of central planning in a city that happened in the middle of the $19^{th}$ century in an effort to transform Paris and to improve traffic flow, navigability and hygiene (see~\cite{Barthelemy_2013} and~\cite{Jordan_1995} for historical details). 

In Fig.~\ref{fig:paris_evo}A we show five instances of the street network (1790, 1836, 1849, 1888, 1999), corresponding to the region clipped to 1790. Highlighted in red are the high BC nodes that lie in the $90^{th}$ percentile. The spatial pattern of the nodes remains virtually identical (with a radial, spoke-like appearance) until 1849, and experience an abrupt change to a ring-like pattern in 1888 which continues to persist for a hundred years. This change in pattern corresponds to the period after the Haussmann transformation, which involved the creation of a number of new roads, broader avenues, new city squares among other things. Yet, it is important to note, that relative to the spatial extent of the region these high BC nodes continue to be located near the city center. Also of note is the relative stability of the edge-density ($\rho_e \approx 0.5$) across the temporal period, reflecting the fact that both nodes and edges are growing at the same rate (Fig.~\ref{sifig:paris_network_evolution}).

The rescaled BC distribution, $\tilde{g_B}$, is identical for all 5 snapshots as seen in Fig.~\ref{fig:paris_evo}C despite the structural changes brought about by the Haussman transformation (this is further indication of the marginal effect of local topological variations in the global BC distribution). In Figs.~\ref{fig:paris_evo}C and D, we show the clustering $\langle C_{90} \rangle$and anisotropy metrics $\langle A_{90} \rangle$ for the different eras, which capture the transition from the radial to the ring pattern, but are nevertheless relatively flat in correspondence with what one would expect to see in the random graph for fixed $\rho_e$. For purposes of comparison, we also plot the averaged metrics for hundred random realizations of each of the five networks that show a remarkable similarity between the original and randomized cities. (The random graphs were generated the same way as in Fig.~\ref{fig:fig3}, i.e we first generate the DT and remove edges until we match the empirical number.) To track the evolution of the BC at the local level, we identify those intersections that are present throughout the temporal interval (within a resolution of fifty meters) and compute their betweenness in each instance of the network normalizing by $N^2$ to provide a consistent comparison, given the historical increase in intersections and roads. In Fig.~\ref{fig:paris_evo}E we plot the temporal evolution of $g_B/N^2$ for these intersections, coloring the points according to their corresponding relative rank. While one observes significant fluctuations in the BC at the local level (as expected), the high BC nodes are relatively stable from 1790-1849. 

After the Haussmann intervention, one observes a dramatic drop in rank of the high BC nodes---corresponding to the ``decongesting" spatial transition from a radial to a circular pattern seen in panel A---after which once again the high BC nodes are relatively stable till 1999.  Yet it is important to note that the load is simply redistributed to a different part of the network, as can be seen by the transition of the middle-ranked nodes to the top positions in the same periods. Furthermore, as indicated by the spatial layout of these ``new" high BC nodes, they continue to be relatively close to the center (few or none are near the periphery), a pattern that is consistent with what one would expect to find for the corresponding random graphs.  

%\section{Discussion}
\vspace{5px}
\noindent {\large{\bf Discussion}}
\vspace{5px}

Taken together our results shed new light on the understanding of structural flow in spatial networks that appear in so many instances, from biological to infrastructure networks. The remarkable invariance in the BC distribution in random planar graphs seems to be a function of the strong constraint imposed by planarity, leaving only the number of nodes $N$ and the number of edges $e$ as tunable parameters vis-a-vis the BC on the network---a markedly different phenomena than seen for non-planar networks, where betweenness is strongly correlated with degree (or local topology). Empirical studies on street networks, analytical calculations on Cayley trees, coupled with simulations of random planar graph models, suggest this to be a consequence of a bimodal regime consisting of a tree-like structure with a tightly peaked branching ratio comprising the high betweenness ``backbone" of the network, and a low betweenness regime dominated by the presence of loops. The transition of nodes between regimes is driven by increasing the density of edges in the network, which has the additional effect of introducing a spatial correlation in the high BC nodes---from being dominated by topology in the low density regime to being strongly dependent on spatial location in the high density regime, features also seen in the spatial distribution of the BC in real cities.  Given that the number of roads and intersection in our sampled cities vary over three orders of magnitude, the similarity in the BC distribution can be explained as a function of the observed narrow range of $\rho_e$.  Indeed, it appears that the characteristics of flow across cities is better characterized by the spatial distribution of the high BC set, as well as the specific location of nodes that lie on this set, rather than global-level statistics.  

On the other hand, the relative lack of sensitivity of the BC distribution to changes in the spatial layout, including distances and local topological variations, has some interesting implications for urban planning. While the random graph models are closer in spirit to so-called self-organized cities that grow organically, the observed evolution of Paris suggests that central planning may also have its limitations. The invariance of the BC distribution suggests that congestion (in the structural sense) cannot be alleviated, but only redirected to different parts of the city. Indeed, the Haussmann transformation succeeded in doing precisely that by improving the navigability of Paris and decongesting the center. However the high BC backbone continued to be closer to the center than the city periphery, a consequence of the spatial distribution being a function of $\rho_e$. For cities with a higher ratio of roads to intersections, the ``decongestion-space" as it were, is expected to be even more limited.

It must be noted that the BC does have limitations in terms of  predicting real time traffic behavior. In particular, weighting edges based only on Euclidean distance artificially places more demand on shorter streets, although in reality these streets may have lower speed limits and thus receive less travel demand ~\cite{leung2011intra}. There is also the issue of spatially irregular travel demand which is overlooked in the betweenness formulation, as all pairs of nodes are given equal weight in the calculation of the global metric~\cite{kazerani2009can}. Various solutions to this route-sampling issue~\cite{Lee_2017} have been proposed; in particular, there have been studies using alternative versions of betweenness that weight each node pair proportional to its perceived travel demand, obtained via both real dynamic data and/or heuristics depending on the study ~\cite{gao2013understanding,chen2012traffic}. The planarity constraint is also alleviated in many cases with multilevel underpasses, public transportation, etc, although the majority of the network still remains planar. We argue that despite these concerns, the results in this study are flexible enough to suggest that load redistribution will be the primary result of planned traffic intervention given static network structure. In particular we can absorb travel preference, distance, speed limits, and other spatially heterogeneous factors into our edge weights, and the invariance of the BC distribution to edge weight adjustment can be used as evidence for these factors not affecting the global load distribution (Cf. Fig.~\ref{sifig:otherdists}). In addition, the construction of detours and alternative paths can be absorbed into the factors affecting local topology, which also leaves the global BC distribution invariant. 

As an example, in Supplementary Note~\ref{sisec:bc_fastest}, we show the case for route sampling. Here, we consider only those streets that lie on the temporally fastest paths (as measured by speed limits) between all Origin Destination pairs in the city, also available from the OpenStreetMaps (OSM) database~\cite{OpenStreetMap}. It is important to note that these constitute a functional and comparatively much smaller (Cf. Table~\ref{sitab:fastest_routes}) subset of the streets and encode more information (dynamics and route sampling) than merely the spatial structure of the network. Moreover for the majority of the considered cities, the fastest paths and the shortest paths \emph{do not coincide} and are markedly different~\cite{Lee_2017}. In Fig.~\ref{sifig:fastest_routes}A we show this subset (in white) overlaid on the full street network (light red) for a city selected from each of the categories (small, medium and large). Fig.~~\ref{sifig:fastest_routes}B shows the BC distribution for the three categories, once again indicating a clear bimodal regime with the peaks located at $N$, while Fig.~~\ref{sifig:fastest_routes}C shows $\tilde g_b = g_b/N$ with all peaks lined up. Finally, in Fig.~\ref{sifig:fastest_routes}D we show the rescaling of the tail with all points collapsing on to a single curve.  

Generally speaking, the study of high BC nodes is an important endeavor as they represent the bottlenecks in the system. In some sense, they represent a generalization of studying the maximum BC node, that governs the behavior of the system in saturation cases where the traffic exceeds the node-capacity. Our analysis suggests, however, that for planar graphs, one needs to take into account the entire high BC set, since the maximum BC node can easily change due to local variations, yet is guaranteed to lie somewhere along the spanning tree that constitutes the backbone of the network. In this respect, further study of the mechanisms governing the spatial distribution of BC is important. Planar graphs are an important class of networks that include infrastructural systems such as power grids and communication networks, as well as transport networks found in biology and ecology~\cite{bio_planar_nets}. In particular, leaf venation networks, arterial networks, and neural cortical networks rely on tree-like structures  for optimal function~\cite{Tekin_2016}. The lessons from this analysis may well be gainfully employed in these other sectors.

%\appendix

\vspace{5px}
\noindent {\large{\bf Material and Methods}}
\vspace{5px}

%\section{Construction of street networks}
%\label{sec:construct} 
\noindent{\bf Construction of street networks} 

The street networks used in our analysis were constructed from  the OpenStreetMaps (OSM) database~\cite{OpenStreetMap}. For each city we extracted the geospatial data of streets connecting origin-destination pairs within a 30 kilometre radius from the city center (referenced from \url{latlong.net}~\cite{latlong}), corresponding to a rectangular area of approximately $60\times60$ square-kilometers with some variability due to road densities, latitude and topographical variations. The $30$ kilometer radius was chosen to encapsulate both high density urban regions and more suburban regions with fewer, longer streets. Furthermore, the choice of scale negates any (minimal) boundary effects on the calculated distribution of the BC~\cite{Gil_2016, Lion_2017}. The locations of the street-intersections were found using an Rtree data structure for expedited spatial search \cite{rtree}. Lattitude and longitude coordinates were projected onto global distances using the Mercator projection, and adjacent intersections lying along the same roads were adjoined by edges with weights equal to the Euclidean distance between the intersections. The resulting street networks are weighted, undirected planar graphs with intersections as nodes, and edges between these nodes approximating the contour of the street network. Aggregate statistics are shown in Tab.~\ref{tab:data}.

\begin{table*}[t]
\begin{tabular}{|c|c|c|c|c|c|}
\hline
{} &   Nodes $N$ & Edges $e$& Length $\ell$ (km) &  Area $A$ (km\textsuperscript{2})&   Density $\rho$ \\
\hline
mean  &  83528.87 & 130253.05 & 17461.68 &                           4600.08 &    18.02 \\
stdev   &  90335.10 & 143060.21 & 15052.83 &                           1926.00 &    15.43 \\
min   &   3349.00 &   5020.00 &  1793.45 &                            777.07 &     1.00 \\
25\%   &  18925.00 &  28518.00 &  5789.36 &                           3184.32 &     5.35 \\
50\%   &  62451.00 &  95797.00 & 12812.46 &                           4411.81 &    14.98 \\
75\%   & 118712.00 & 178773.00 & 23751.22 &                           5873.67 &    26.59 \\
max   & 612418.00 & 976040.00 & 82586.30 &                          11562.73 &    93.47 \\
\hline
\end{tabular}
\caption{{\bf Aggregate statistics for the 97 street networks.} Shown are the average, standard deviation, minimum, maximum and various percentile values for the area $A$,  number of intersections (nodes) in the network $N$, number of roads (edges) $e$, total length of streets $l$ and the density $\rho = N/A$ of intersections. Details for individual cities shown in Tab. \ref{sitab:networkproperties}.}
\label{tab:data}
\end{table*}

%\section{BC of Cayley trees}
%\label{sec:bc_cayley}

\clearpage
\noindent{\bf BC of Cayley trees}

Let us consider a perfect Cayley tree of size $N$ with fixed branching ratio $k$ and all leaf nodes at the same depth. Adopting the convention $l=L$ for the leaf level and $l = 0$ for the root, a node on the $l$-th level has $k-1$ branches directly below it at the $(l+1)$-th level, each with $M_{l+1}$ children such that the set of branches $\{n_i\}$ stemming from this node will have sizes $\{n_i\}=\{M_{l+1},...,M_{l+1},N-M_{l}\}$. For fixed $k$ there are $k-1$ copies of the term $M_{l+1}$ which is of the form
\begin{equation}
M_{\lambda}=\sum_{l'=0}^{L-\lambda}k^{l'} =\frac{1-k^{L-\lambda+1}}{1-k}.
\label{eq:ml}
\end{equation}
The betweenness value of a vertex $v$ in any tree is given by  $g_B(v)=\sum_{i<j}n_in_j$ where $i$, $j$ are indices running over the branches coming off of $v$ (excluding $v$), and $n_i,n_j$ are the number of nodes in each branch~\cite{bet_examples}. Combining this with Eq.~\ref{eq:ml} gives us the betweenness of $v$ at level $l$ thus
\begin{equation}
g_B(v\vert k,l)
={k-1\choose 2}M_{l+1}^2
+(k-1)M_{l+1}\left(N-M_{l}\right),
\end{equation}
from which it is easy to see that for any level $l$, the betweenness scales as $g_B(v\vert k,l)\sim O(Nk^{L-l})$. Thus, absorbing $k^L$ into the leading constant and letting $g_B(v\vert k,l)\approx ANk^{-l}$, we have that since $g_B$ is completely determined by the level $l$ in which it lies in the tree,
\begin{equation}
P(g_B)=\sum_{l}P(g_B\vert  l)P(l).    
\end{equation}
Now, using the fact that $P(l)=\frac{k^l}{N}$ and $P(g_B\vert  l)=\delta_{g_B,ANk^{-l}}$, we have that
\begin{equation}
P(g_B)=Ag_B^{-1}.
\label{eq:cayley1}
\end{equation}

%\section{Spatial metrics for high BC nodes}
%\label{sec:spatial_bc}
\noindent {\bf Spatial metrics for high BC nodes}

To measure the clustering, we specify a threshold $\theta$---i.e. we isolate nodes with a BC above the $\theta$-th percentile---and then compute their spread about their center of mass, normalizing for comparison across networks of different sizes, thus,
\begin{equation}
C_{\theta}=\frac{1}{N_\theta\langle \textbf{X}\rangle}\sum_{i=1}^{N_\theta}\vert\vert\textbf{x}_i-\textbf{x}_{cm}\vert\vert.
\label{eq:cluster}
\end{equation}
Here $\textbf{x}_{cm}=\sum_{i=1}^{N_\theta}\textbf{x}_i$, $N_\theta$ is the number of high betweenness nodes isolated, $\{\textbf{x}_i\}$ specify their coordinates, and $\langle \textbf{X}\rangle$ is the average distance of \emph{all nodes} in the network to the center of mass of the high BC cluster,
\begin{equation}
\langle\textbf{X}\rangle=\frac{1}{N}\sum_{i=1}^{N}\vert\vert\textbf{x}_i-\textbf{x}_{cm}\vert\vert.
\end{equation}
Eq.~\ref{eq:cluster} quantifies the extent of clustering of the high BC nodes relative to the rest of the nodes in the network, with increased clustering resulting in low values of $C_{\theta}$.  

In order to more precisely quantify the transition between the topological and spatial regimes, a clue is provided by the increasingly isotropic layout of the high BC nodes with increasing edge-density. To measure the extent of this observed (an)isotropy, we define the ratio,
\begin{equation}
A_\theta=\frac{\lambda_1}{\lambda_2},
\label{eq:iso}
\end{equation}
where $\lambda_1 \leq \lambda_2$ are the (positive) eigenvalues of the covariance matrix of the spatial positions of the nodes with BC above threshold $\theta$. The metric is unitless and measures the widths of the spread of points about their principal axes, analogous to the principal moments of inertia. Low values of $A_{\theta}$ correspond to a quasi one-dimensional structure with large anisotropy, whereas the system becomes increasingly isotropic for larger values until it is roughly two-dimensional as $A_\theta \rightarrow 1$. 

The detour factor measures the average extent to which paths between two locations deviate from their geodesic distance and is given by
\begin{equation}
D=\frac{1}{N(N-1)}\sum_{i\neq j}\frac{d_G(i,j)}{d_E(i,j)}.
\label{eq:detour}
\end{equation}
Here $d_E(i,j)$ is the euclidean distance between nodes $i,j$ and $d_G(i,j)$ is their distance-weighted shortest path in the network $G$. 

%\section{Distance dependence of BC}
%\label{sec:distance_bc}
\noindent{Distance dependence of BC}

In our simulations, nodes were located on a $100 \times 100$ grid with coordinates in $\mathbb{R}^{2} \in [-50,50]$. The center of the grid was chosen as the origin $(0,0)$ and the average betweenness $\langle g_B(r)\rangle$ is computed over all nodes that are located at a distance $r$ from the origin, advancing in units of $r=1$, until  we reach the grid boundary $r=50$. In order to restrict $\langle g_B(r)\rangle$ to the interval $[0,1]$ we measure the rescaled quantity 
\begin{equation}
\langle g^{\star}_b(r)\rangle = \frac{\langle g_{B}(r)\rangle - \min\langle g_B(r)\rangle}{\max\langle g_B(r)\rangle)- \min \langle g_{B}(r)\rangle},
\end{equation}
for different values of $\rho_e$. This was done to compare our results  to the corresponding expression in random geometric graphs, which was analytically calculated for (the somewhat artificial) limit of an infinitely dense disk of radius $R$~\cite{Giles_2015}.

\begin{acknowledgments}

This work was partially supported by the US Army Research Office under Agreement Number W911NF-17-1-0127. MB thanks the city of Paris (Paris 2030)
for funding and the geohistoricaldata group for discussions and data.

\end{acknowledgments}

\bibliographystyle{naturemag}
\bibliography{refs}

\end{document}